# A New PLS-II In-Vacuum Undulator and Characterization of Undulator Radiation

D-E. Kim, H-H. Lee, K-H. Park, H-S. Seo, T. Ha, Y-G. Jeong, H-S. Han, W.W. Lee,

J-Y. Huang, S. Nam, K-R. Kim and S. Shin

*Pohang Accelerator Laboratory, POSTECH, Pohang, 790-784, KOREA*

This paper describes the result of overall studies from development to characterization of undulator radiation. After three years of upgrading, PLS-II [1, 2] has been operating successfully since 21st March 2012. During the upgrade, we developed and installed an in-vacuum undulator (IVU) that generates brilliant X-ray beam. The IVU with 3 GeV electron beam generates undulator radiation up to ~ 21 keV using 11th higher harmonic. The characterizations of the undulator radiation at an X-ray beam line in PLS-II agreed well with the simulation. Based on this performance demonstration, the in-vacuum undulator is successfully operating at PLS-II.

## I. INTRODUCTION

Newer generation light sources with medium electron energy around 3 GeV as well as SASE-FEL are adopting in-vacuum undulator (IVU) to obtain short-wavelength and high-brilliance X-rays from undulator with short magnet period and required magnetic field. Pohang Accelerator Laboratory (PAL) [3] had developed IVU during PLS-II project. With operating experience of in-vacuum revolver undulator from SPring-8 [4] and conventional IVU from ADC [5] at the PLS [6], the new IVU with 20 mm magnetic period and 1.8 m long undulator length was developed and have been operated in PLS-II.

After completing design, field shimming and assembly of IVU, we had successfully installed and operated IVU in the PLS-II storage ring. During commissioning of PLS-II, the undulator radiation was characterized at the X-ray scattering beam line in PLS-II. As a result, up to 21 keV undulator radiation using $11^{th}$ harmonic is obtained and available at beamline. Now total 10 IVUs have been operated to generate short wave-length high brilliance X-rays for X-ray scattering, diffraction and nano-imaging studies in PLS-II.

In this paper we described the results of overall studies from development to X-ray characterization from undulator. Section 2 introduces overall works to develop IVU. Section 3 describes the investigation on their effects on stored electron beam. The characterization of the IVU radiation at the X-ray scattering beam line is described in Section 4. Finally, Section 5 presents summary and conclusive remarks.

## II. DEVELOPMENT OF IN-VACUUM UNDULATOR

A schematic drawing of the IVU is shown in Fig.1 (a) and main parameters of the IVU are introduced in Table 1. The IVU consists mainly of mechanical supporter, magnet array and vacuum system. The IVU is equipped with online built-in magnetic measurement for the end regions to assess the accuracy of the assembly and degradation coming from the radiation damage or high temperature.

### 1. Undulator design

The designed IVU is a conventional hybrid type undulator, which utilizes ferromagnetic poles to concentrate the magnetic fluxes. The advantage of the hybrid type undulator is that its effective peak field is stronger for a same period than that of the pure permanent magnet type. It is also known that the performance of a hybrid type insertion device depends on the mechanical accuracy of the device rather than on the material properties of the magnet that are more difficult to control than the mechanical accuracy. The non-uniform magnetization of the permanent magnets has been observed to affect the performance of the undulator, particularly, when the gap is small and the electron beam is very close to the magnet. It should be also noted that this kind of local non uniformity does not show up in the Helmholtz coil measurement, which measures only the total dipole moment of the block.

$Sm_2Co_{17}$ type rare-earth magnet is selected for the permanent magnet. The remanence of $Sm_2Co_{17}$ ranges between1.12 T and 1.20 T with a coercivity of 9.8 kOe ~11.4 kOe, which is about 10% inferior to the state of the art NdFeB class magnet. Although $Sm_2Co_{17}$ is more brittle and difficult to handle than the NdFeB, it has a big advantage in case of a long term radiation damage by high energy electron beam irradiation compared to the NdFeB based magnet [7]. Since top-up injection is a default operation mode in PLS-II, the beam injection at the minimum operation gap is anticipated and the long-term stability against radiation damage is very important. In addition, $Sm_2Co_{17}$ has a higher Curie temperature (about 660 ℃), and a working temperature with small coefficient of -0.03 %/K (-0.11~ -0.12 %/K for NdFeB). Therefore, it is very safe from overheating caused by accidental heating during the vacuum baking. Moreover, the effect of temperature rise coming from the image current heating is minimized. Of course, the magnetic structure is cooled by the liquid cooling water system (LCW) during baking to protect the magnet structure from overheating. The LCW cooling is also aiming to remove any drift in the spectrum coming from the temperature changes in the magnet structure from the image current heating. Adoption of Sm2Co17 improves the drift of the spectrum from the small temperature drift during the cooling and current cycles. Two coating methods are tested for out-gassing: ion vaper deposit (IVD) coating and Ni coating. The IVD coating was un-acceptable, on the other hand, Ni-coating was good enough for the purpose of IVU. The undulator was assembled using Ni coating, and there has been no vacuum problems caused by the surface outgassing.

The magnetic structure is designed by using RADIA [8], which was developed at ESRF for insertion device calculation, and cross checked by using ANSYS 3D and OPERA. Due to the limited computer resources, only 10 periods of the whole undulator are modelled to assess the periodic part and transition parts. The RADIA model is shown in Fig. 1 (b). The magnet size chosen is 65 mm (W) × 27 mm (H) × 7.0 mm (T) with 4.0 mm chamfering at the edges. The pole dimension is 40 mm (W) × 20 mm (H) × 3.0 mm (T). Although the optimum effective peak field was achieved using thinner pole thickness, 3.0 mm thickness is chosen due to the mechanical consideration. The difference between the optimum and the current configuration was small (about 3%). The longitudinal distribution of the blocks is such that the field is anti-symmetric in $z=0$, which is the undulator center in the beam direction. This anti-symmetric configuration has a benefit of systematic zero $1^{st}$ field integral for all gaps for ideal materials and geometry. Any deviation from zero integral is attributed to the imperfections in geometry or magnetic field, which can be easily tuned to be zero. The weakness of this anti-symmetric configuration is that the $2^{nd}$ field integral is not systematically zero. However, the $2^{nd}$ integral requirement is weak and it can be minimized while designing the transition parts. The transition sequence at the ends is determined by adjusting the space to achieve the periodic field and minimize the $2^{nd}$ orbit displacement at the end of the undulator. The calculated effective field at nominal gap of 6.0 mm is 0.815 T. The magnetic length of the undulator is limited to about 1400 mm, resulting in a flange to flange distance of 1800 mm, which is near the maximum installable device length for the short straight section of PLS-II.

2. **Field measurement and shimming**

The assembled undulator is measured and corrected using 3-axis hall probe bench. The longitudinal position is measured using a laser interferometer with an accuracy better than 1 μm. The flatness of the bench is calibrated using laser tracker to within 20 μm. The 3-axis hall probe is calibrated for angular error, nonlinearity, and planar hall effects. The temperature of the entire measurement room is controlled within ±1 K. Also a flip coil system is used to calibrate the planar hall effect by measuring the accurate residual field integral. The accuracy of the flip coil system was ± 5 G·cm.

The undulator is corrected before installing the vacuum chamber. The reproducibility of the undulator assembly is checked by re-measuring the undulator after dis-assembling the vacuum chamber several times, and it is concluded that the assembly is reproducible enough.

Pole shimming is carried out by adjusting the pole heights by using copper shims of 10 μm, 30 μm, 50 μm, 100 μm and combinations of the shims. Pole tuning was conducted in 3 modes as depicted in Fig. 2. Fig. 2(a) shows shimming that can correct the skew dipole components and can also correct $y$-orbits by tilting the vertical magnetic field and by creating the horizontal field. For 3 mrad tilting of a set of upper and lower pole, we can expect about 15.6 G·cm skew dipole corrections. The actual measurement agrees with the calculation within 20%. These corrections at several points along the undulator can straighten the $x$-orbit, which is needed to improve the optical phase errors. Fig. 2(b) shows the correction schemes for normal quadrupole components. By canting a pole, we can adjust the transverse gradient of the residual field integral. For our case of 3 mrad canting of an upper and lower pole, we can change the integrated normal quadrupole component by 30 G. Fig. 2(c) shows the correction scheme for the normal dipole component and the horizontal orbit. This kind of shimming can correct the field amplitude and orbit straightness. The correction is mostly used to minimize the optical phase jitters. After the pole shimming, the optical phase advance per pole becomes uniform, and the average orbit straight with minimal optical phase errors. The requirement for optical phase error excluding transition parts is 5 degree in rms value. The final optical phase errors of the undulators were about 3.5 degrees in rms value, which satisfies the requirements. Also the ideal spectrum with zero beam emittance for the ideal field and the measured field is estimated and compared using B2E code. The calculated spectrum based on the measurement data has achieved 95% of the ideal spectrum at 5-th harmonic. Fig. 3 shows one example of optical phase errors after final pole tuning, and in Fig 4, the estimated spectrum based on the measured field, and the ideal spectrum based on the ideal field is compared. The figure shows that the estimated radiation from the undulator is very close to the ideal one. Here we considered the $5^{th}$ harmonic at 6.0 mm gap with 3.0 GeV electron beam in the calculation.

The pole tuning described optimized the integrated normal dipole, skew dipole, normal quadrupole components, and the next few low order integrated multipole errors were corrected using arrays of trim magnets. The smallest unit is a 1.5 mm diameter and 2.0 mm thick cylindrical shape. Compensation of the field profile is calculated by using RADIA, and the trim magnets are assembled at the end of the undulator. There are three rows in $z$ for each side of trim magnets. One row is used to correct the normal components and the second row is used to correct skew components. The third row is used for reserve or fine tuning. Using this trim magnet system, the integrated $B_x$ and $B_y$ profile could be controlled easily and predictably.

### III. COMMISSIONING AND OPERATION OF THE IVU

The operation of the IVU in the PLS-II storage ring has brought about several effects that can degrade the overall beam performance. The effects of IVU on the beam are mainly changes in beam parameter, reduction of dynamics aperture, residual orbit distortion, and vacuum increase in the storage ring. During the PLS-II commissioning, we investigated the effects of the IVU on the stored beam in the storage ring.

In storage rings of the recent third-generation light source, insertion devices are placed in the region that has finite dispersion, thus affecting the beam parameters such as emittance and energy spread [9]. The beam-parameter changes are listed in Table 2. Here the vertical tune shift and beta-function distortion are caused by the orthogonal component of magnetic field vertical to the trajectory. The changes of emittance, energy spread, and energy loss are due to undulator radiation. These changes by IVU are less than a few percent, since the peak field and period length of the undulator are small [10].

The origin of the nonlinearities of the insertion device effects can be critical. The periodic beam path through the insertion device causes nonlinear and periodic magnetic interaction. The effect of the nonlinear intrinsic and transverse field roll-off from the IVU on particle tracking is very small and negligible due to the small peak field and the period length [10]. However, unlike an ideal insertion device, the first and second field integrals of the insertion device distort the equilibrium orbit in a real insertion device. Measured rms orbit distortions by IVU in the PLS-II are 1 to 20 $\mu m$ for the horizontal and 4 to 50 $\mu m$ for the vertical plane. In order to compensate the orbit distortion by the kick from IVU, the auxiliary steering magnet for insertion-device feed-forward and global orbit correction [11] has been introduced, which suppresses the orbit distortion significantly better than a factor of ten as shown in Fig. 5.

## IV. CHARACTERIZATION OF THE IVU RADIATION

### 1. Theoretical description

The main purpose of the insertion device is to generate intense synchrotron radiation in the electron storage ring. The spectral brightness for a bending magnet and the IVU in PLS-II is shown in Fig. 6. In the figure, the spectral brightness of the 1st, 3rd, 5th and 7th harmonics along the IVU's gap is introduced together with the brightness from the bending magnet. Compared with the radiation from the bending magnet, the spectral brightness of the IVU radiation is 1000 times higher.

The maximum angular deflection of the electron beam in the undulator is much smaller than the opening angle of the radiation cone, and the wavelength radiated from insertion device is

$$\lambda_n(\theta) = \frac{\lambda_w}{2n\gamma^2}(1+\frac{K^2}{2}+\gamma^2\theta^2), \tag{1}$$

where $K$ is the dimensionless field strength parameter, $n = 1,2,3, \ldots$ is harmonic number, $\lambda_w$ is the undulator magnet period, $\gamma$ is the relativistic parameter, and $\theta$ is radiation observation angle. The spectral width of the peak emission along beam axis ($\theta = 0$) is given by

$$\frac{\Delta\lambda}{\lambda} = \frac{1}{nN}, \tag{2}$$

where $N$ is the number of undulator periods. The line width of the undulator is also affected by various effects including the field error and source size. Source size and divergence of the radiation from an ideal undulator are given by

$$\sigma_r = \frac{\sqrt{\lambda L}}{2\pi}, \sigma'_r = \sqrt{\frac{\lambda}{L}}, \tag{3}$$

where $L$ is undulator length. For a realistic synchrotron light source, the finite beam emittance of the particle beam must be taken into account as well often even when the dominant emittance beam is larger than the diffraction limited photon beam emittance. In Fig. 7 (a), both undulator spectra from zero emittance and finite emittance electron beams are calculated by using the SPECTRA program. Here IVU parameters with $K = 1.59$ at $\theta = 0$ are used in the calculation. While the spectrum from zero emittance electron beam demonstrates the line character of undulator radiation well, the undulator radiation from the finite emittance electron beam shows a broad continuous spectrum, including even harmonics due to the spatial effects of undulator radiation. The overall spatial intensity distributions are shown in Fig. 7 (b) and (c), including a complex set of different radiation lobes depending on frequency, emission angle, and polarization.

2. **Measurement of the IVU radiation**

5A XRS BL (X-ray Scattering Beamline) in PLS-II is dedicated to materials science research. The main scientific programs are in-situ as well as ex-situ x-ray scattering experiments on thin films, soft materials, and nano-structured materials. The optical layout of 5A XRS BL is shown in Fig. 8. The optical components consist of a double crystal monochromator comprised of a pair of Si (111) crystals, a vertical focusing mirror, and a horizontal focusing mirror.

Figure 9 shows the spectrum from the IVU installed for the 5A XRS beamline. The spectrum was measured with the photocurrent from the Cu-plate located after Si (111) double crystal monochromator (DCM). As shown in the figure, the spectra are clear to the 11$^{th}$ harmonics. Due to the limitation of the energy scan range of the double crystal monochromator, we cannot measure harmonics higher than 11th. Figures 9 (b) and (c) present the images of the harmonics taken at a distance of 20 m from the source with the gap of 6.97 mm. Compared with the simulated images in Fig. 8 (b) and (c), the measured images indicate the central cone very clearly while the other structures are symmetric. This directly ensures that the devices are well developed as they were designed.

## V. SUMMARY

In this paper, we introduced a new IVU designed and developed for an X-ray source at PLS-II. Their effects on the beam operation had been also described and its effect on beam operation was negligible or suppressible. The characterizations of the undulator radiation at an X-ray beam line in PLS-II agreed well with the simulation. Based on this performance demonstration, the IVU is successfully operating at PLS-II.

## ACKNOWLEDGEMENT

We would like to thank H. Wiedemann for useful discussions and guidance. This research was supported by the Converging Research Center Program through the Ministry of Science, ICT and Future Planning, Korea (NRF-2014M3C 1A8048817) and (NRF-2015R1D1A1A01060049).

Table 1. Main parameters of the PLS-II in-vacuum undulator.

| Parameter | Value | Unit |
|---|---|---|
| Undulator type | Hybrid | |
| Longitudinal Symmetry | Anti- Symmetric | |
| Period length | 20 | mm |
| Working gap | 5.0~16 | mm |
| Effective field | 0.815 | T |
| Magnetic length | 1400 | mm |
| Flange to flange length | 1800 | mm |
| Period number | 67 | |
| Magnet material | $Sm_2Co_{17}$ | |
| Phase error | < 5 | Degree |
| Gap straightness | < 1.0 | μm |
| Vacuum pressure | < 3E-9 | Torr |

Table 2. Main beam parameter changes by the IVU.

| Parameter | Value | Unit |
|---|---|---|
| $\Delta Q_y$ | 0.002 | |
| $\Delta \beta_y / \beta_y$ | 1.04 | % |
| $\Delta \varepsilon$ | 0.04 | nm rad |
| $(U - U_0)/U_0$ | 0.47 | % |
| $(\sigma_E - \sigma_E^0)/\sigma_E^0$ | -0.12 | % |

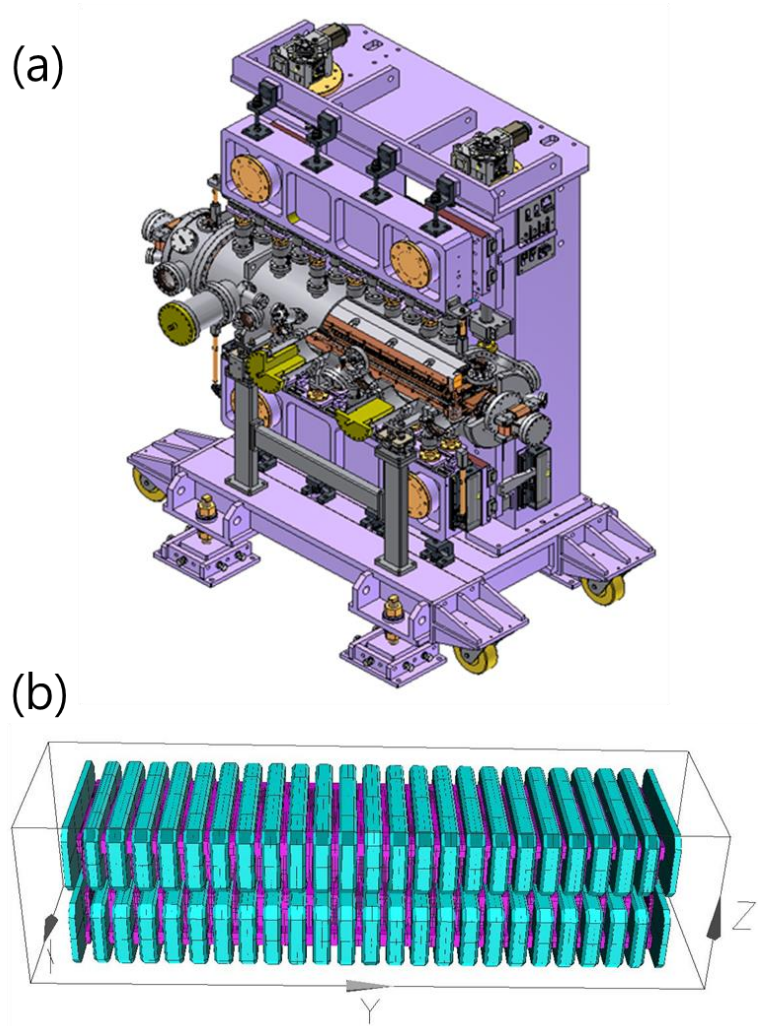

Fig. 1. (a) Schematic Drawing of the IVU. The IVU consists mainly of mechanical supporter, magnet array and vacuum system. (b) The RADIA model. The magnet size chosen is 65 mm (W) × 27 mm (H) × 7.0 mm (T) with 4.0 mm chamfering at the edges. The pole dimension is 40 mm (W) × 20 mm (H) × 3.0 mm (T).

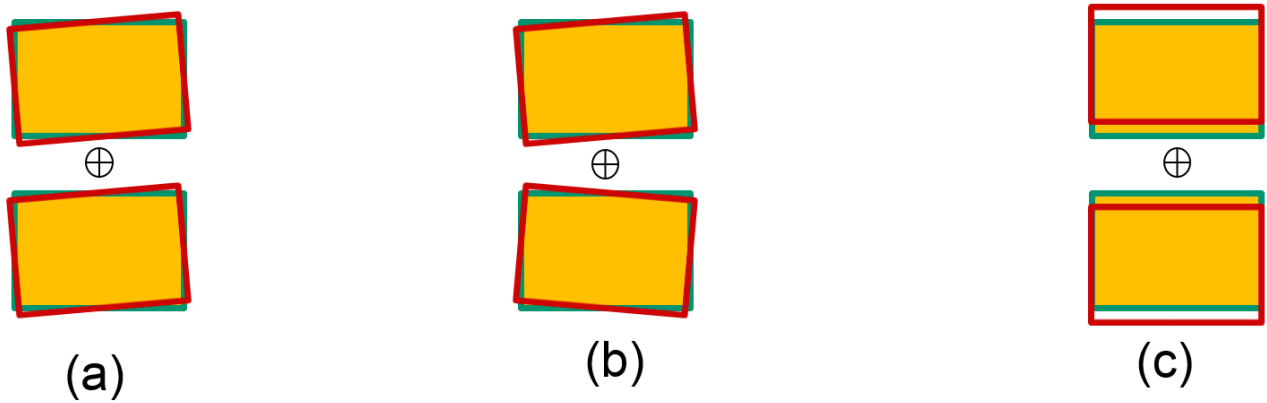

Fig. 2. Schematics of pole tuning. (a) This kind of tilting of a pole is used to correct the skew dipole component. (b) This canting of a pole is used to correct the normal quad. Component. (c) This pole shimming is used to correct the normal dipole component.

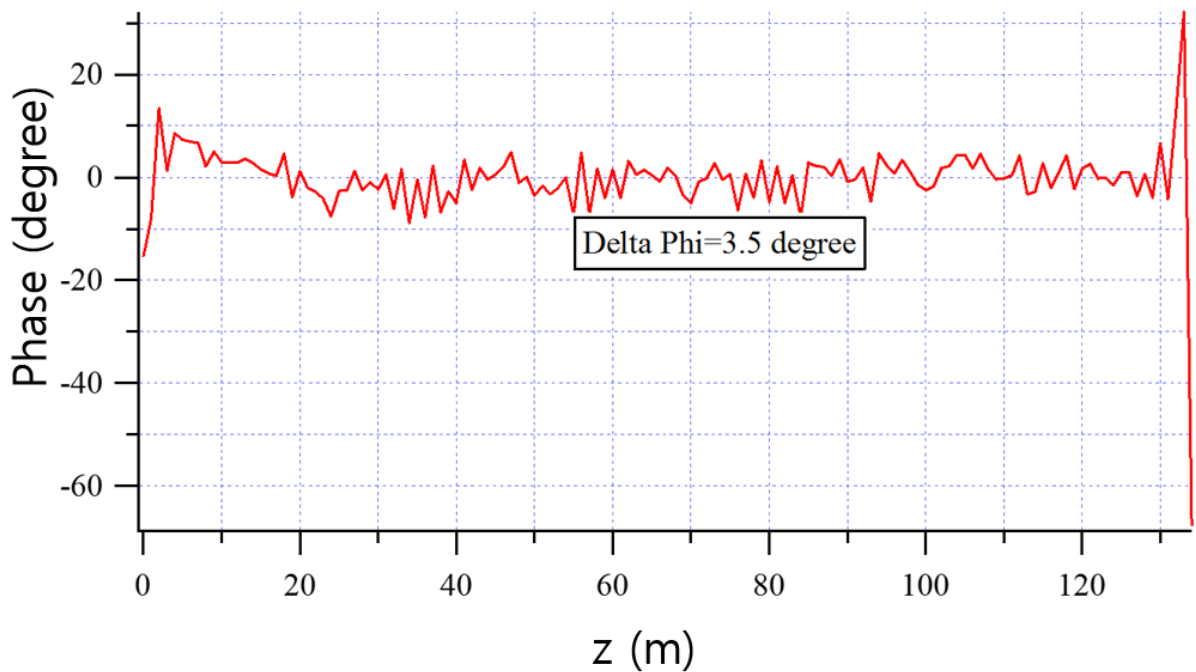

Fig. 3. Optical phase errors after final pole tuning. The requirement for optical phase error excluding transition parts is rms 5 degree.

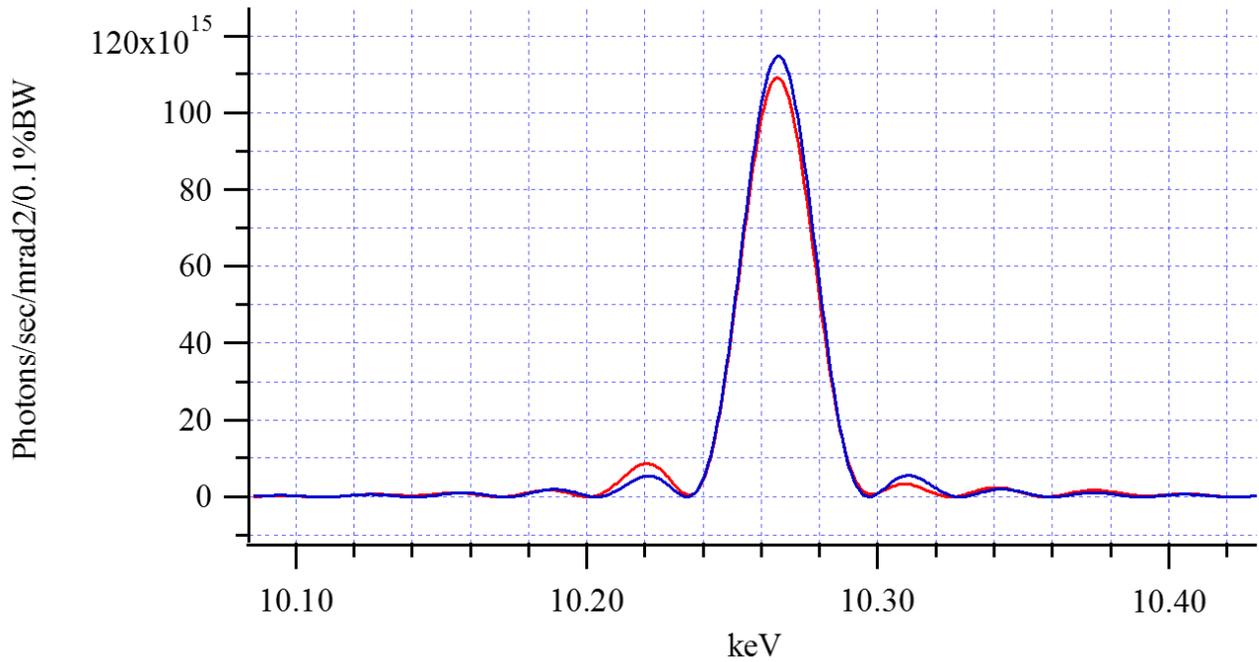

Fig. 4. Spectrums from ideal field (blue line) and measured field (red line). The figure shows that the estimated radiation from the undulator is very close to the ideal one. Here the 5th harmonic at 6.0 mm gap with 3.0 GeV electron beam is considered in the calculation.

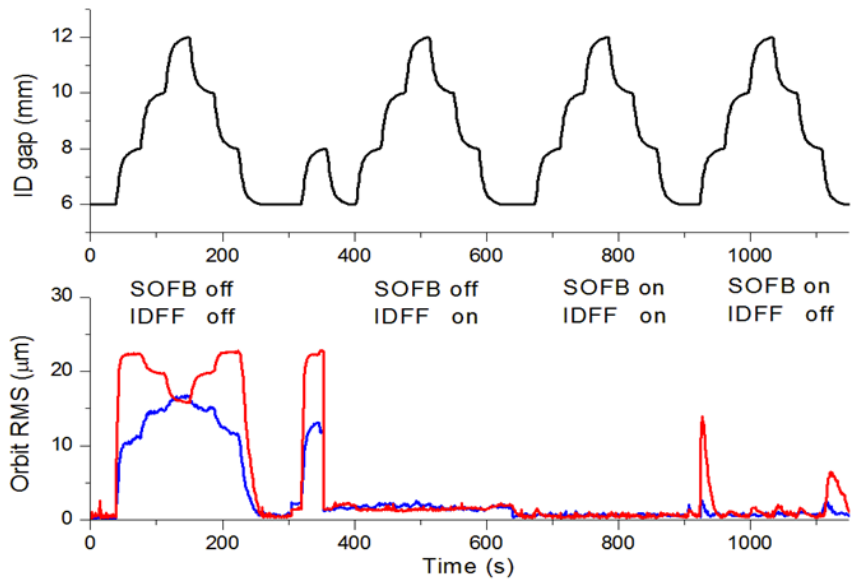

Fig. 5. Orbit distortion due to one IVU gap change. ID feed-forward and global orbit correction systems suppress orbit distortion significantly.
14

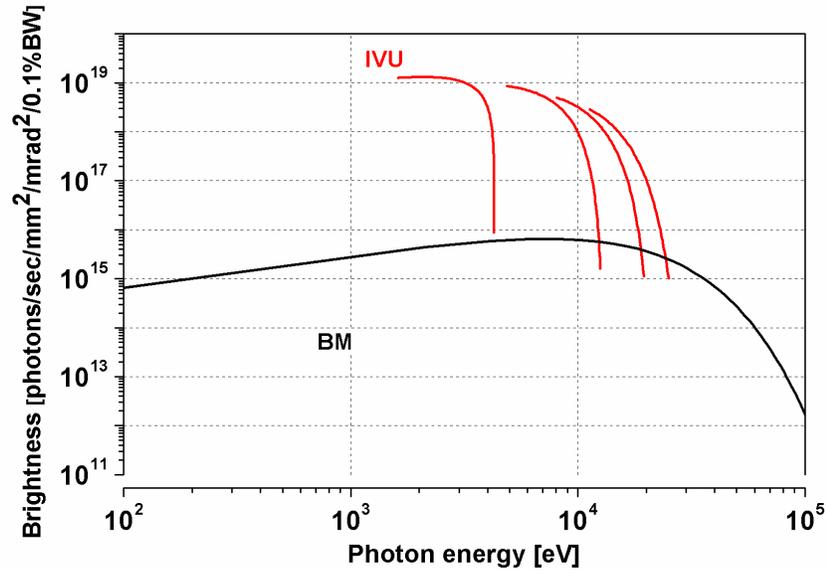

Fig. 6. Spectrum (in-vacuum undulator vs bending magnet). The spectral brightness of the 1st, 3rd, 5th and 7th harmonics along IVU's gap are introduced together with the brightness from the bending magnet.

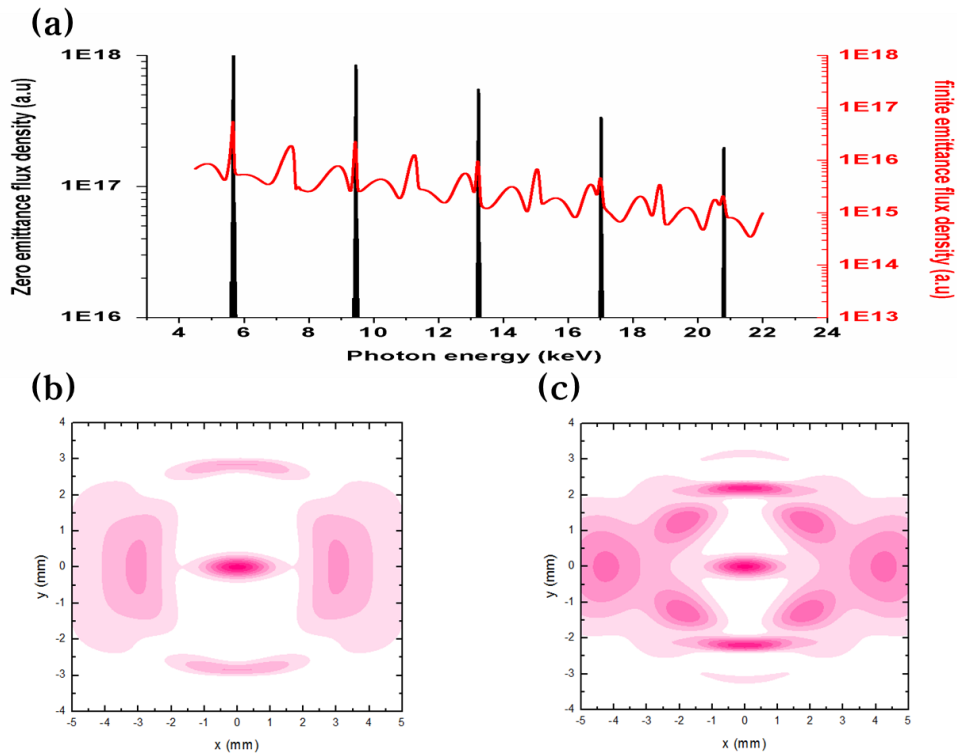

Fig. 7. (a) Undulator spectrum of the PLS-II calculated at the K = 1.58. (b) and (c) are the calculated undulator beam images of the 3rd and 5th harmonics, respectively.

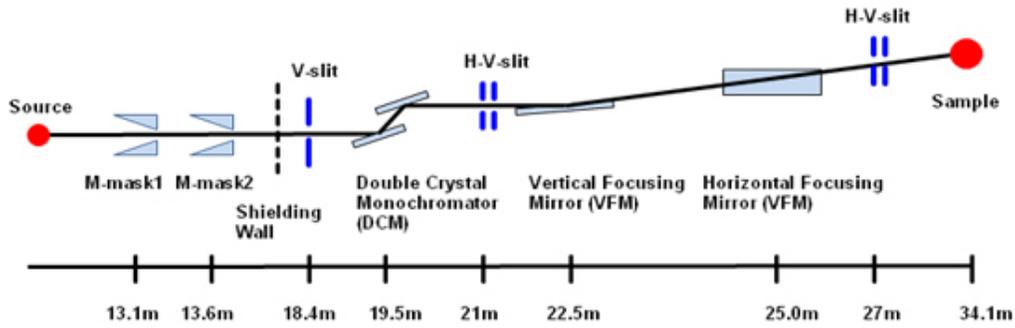

Fig. 8. Beamline layout. Optical components of beamline consist of a double crystal monochromator, vertical focusing mirror, and a horizontal focusing mirror.

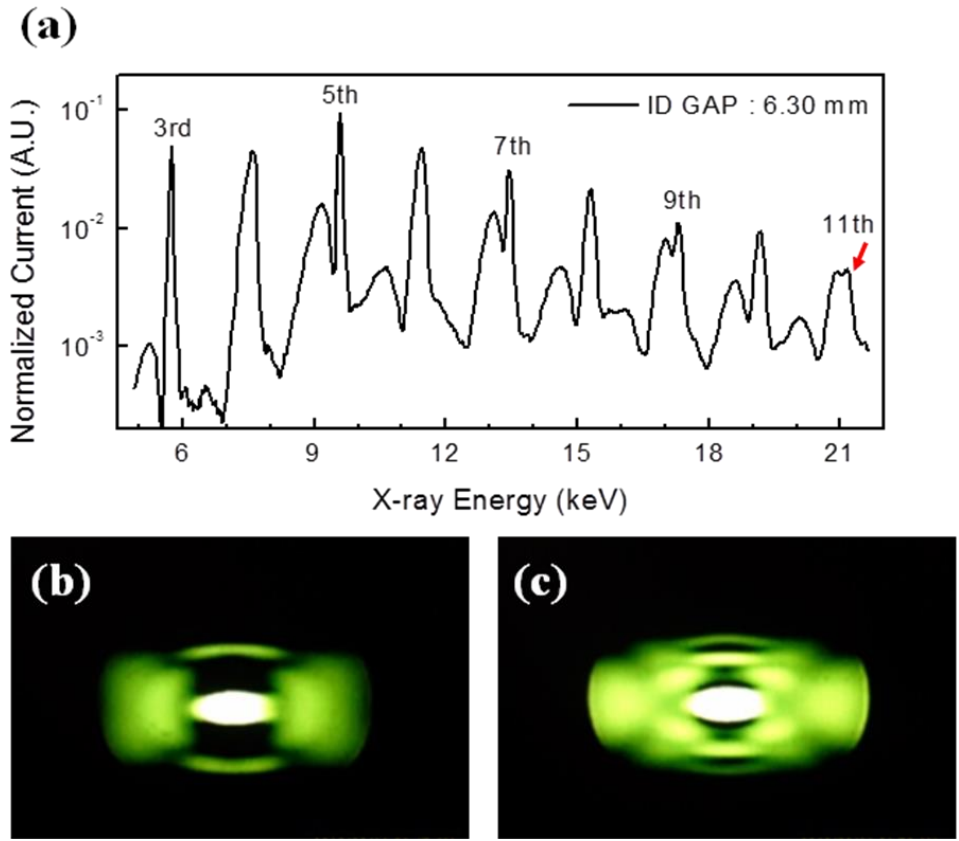

Fig. 9. (a) Undulator spectrum of the PLS-II measured at the gap of 6.3 mm. (b) and (c) are the undulator beam images of the 3rd and 5th harmonics, respectively.

16